\documentstyle[11pt,paspconf]{article}

\begin{document}

\title{Star Formation in High Redshift Galaxies}

\author{Jill Bechtold}
\affil{Steward Observatory, University of Arizona,
    Tucson AZ 85721}

\author{Richard Elston}
\affil{Department of Astronomy, University of Florida, Gainesville FL 32611}

\author{Howard K. C. Yee}
\affil{Department of Astronomy, University of Toronto, Toronto, Ontario M5S 3H8, Canada}

\author{Erica Ellingson}
\affil{Center for Astrophysics and Space Astronomy, University of Colorado, Boulder CO 80309}

\author{Roc M. Cutri}
\affil{IPAC -- Caltech, Pasadena CA 91125}


\begin{abstract}
Observations of the high redshift Universe, interpreted in
the context of a new generation of computer simulated
model Universes,  are providing new insights into the processes
by which galaxies and quasars form and evolve, as well as the
relationship between the formation of virialized, star-forming
systems and the evolution of the intergalactic medium.  We describe
our recent measurements of the star-formation rates, stellar 
populations, and structure of galaxies and 
protogalactic fragments at $z\sim$2.5, including narrow-band imaging in 
the near-IR, IR spectroscopy, and deep imaging from the ground and 
from space, using $HST$ and $ISO$.

\end{abstract}



\section{Introduction}

Observations of the Hubble Deep Field, and other surveys of high
redshift galaxies described in these proceedings, are contributing
to a new picture of how large galaxies such as the Milky Way were 
assembled.  One interpretation of the data so far is that large galaxies
collapse out of what appears as several star-forming proto-galactic 
fragments at $z\sim 2-3$ (Pascarelle et al. 1996; Haehnelt, Steinmetz $\&$ 
Rauch 1998).
Here we describe our search for H$\alpha$ emission from the star-forming
regions in high redshift galaxies, which is redshifted into the near-infrared. 
We focus on one 
of the outstanding questions about these objects: is the
burst of star-formation  seen in the UV continuum producing the dominant
stellar population (by mass)?  We show that for most high redshift objects, 
the data available to date provide only weak constraints on the
age of the dominant stellar population (see also the discussion by
Sawicki $\&$ Yee 1998).  An exception is our observation
of the observed mid-IR (rest-frame near-IR) continuum of the
$z=$2.7 galaxy MS1512-cB58 using $ISOCAM$.  For this object, we
can put a strong  
limit on the fraction of the mass comprised of an old population of stars.  

\section{The $z=$2.515 gravitational arc of Abell 2218}

To illustrate the ambiguity in the age of the dominant stellar population,
in this section we describe observations of the $z=$2.515 galaxy 
which is lensed by Abell 2218.
Abell 2218 is a rich cluster of galaxies at $z=0.171$ (Abell et al. 1989;
Kristian et al. 1978). It was one of the first to
to be observed to have arcs caused by the lensing of background
galaxies (Lynds $\&$ Petrosian 1986; Pello-Descayre et al. 1988),
and has been the subject of many studies.  Of particular interest
is the discovery  by Ebbels and collaborators (E96)
that one of the brightest blue arcs is a star-forming
galaxy at $z=$2.515 (Arc A, or galaxy $\#$384).
The appearance of the optical spectrum of the arc,
as well as the optical/IR spectral energy distribution, imply
that Arc A is typical of the star-forming galaxies
discovered by Lyman dropout techniques (Steidel et al. 1996a)
or in the Hubble Deep Field (Williams et al. 1996, Lowenthal et al. 1997,
Cohen et al 1996, Steidel et al. 1996b), except that its apparent magnitude
is boosted by about a factor of 14.5 by lensing (Saranti, Petrosian
$\&$ Lynds 1996).   Arc A is
similar in apparent properties to MS1512-cB58,
described below.

\begin{figure}[ht]
\plotfiddle{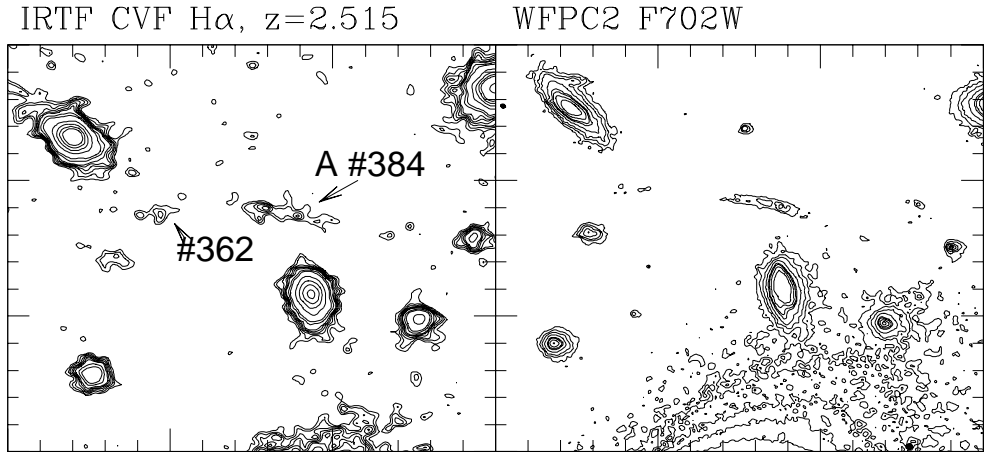}{3.0truein}{0}{90}{90}{-275}{-800}
\caption{Contour plots of H$\alpha$ image of Arc A ($\#$384) and
Arc $\#$362 (left)
at the same scale and orientation as the WFPC-2 image (right).
North is up and east is to the left; 30$\arcsec$ $\times$ 30$\arcsec$
is shown.
Arc A and the optically faint Arc $\#$362,  both appear to have excess
emission in the narrow band image compared to the K image.
Arc $\#$362 appears to be as blue as Arc A, compared to the
numerous arcs at lower redshift which are red (e.g. plate 1 of Saraniti,
Petrosian $\&$ Lynds 1996).  Thus Arc $\#$362 is probably
a counterimage of Arc A, at $z=2.515$.  Another spectroscopically
confirmed counter-image of Arc A, $\#$468 (E96) was not in our
field of view.
}
\end{figure}

We obtained a narrow-band image of
H$\alpha$ emission from A2218's Arc A using the IRTF NSFCAM
and tunable narrow-band filter, or CVF (R$\sim$90, Figure 1).
Further details are provided in Bechtold $\&$ Elston (1998, in preparation).
For Arc A, the
excess emission in the CVF image compared to the $K'$ image
implies  a rest-frame equivalent width of $\sim$85 $\AA$ and
a narrow band flux of 7.35 $\times$ 10$^{-16}$
ergs cm$^{-2}$ sec$^{-1}$.
For $z$ = 2.515,  this corresponds to
a luminosity in the excess narrow-band emission
of 3.25$\times$10$^{43}$ $h_{75}$$^{-2}$ ergs sec$^{-1}$ for
$q_0$ = 0.1. 
The observed H$\alpha$ luminosity implies
a total observed star-formation rate (SFR) of
290 $h_{75}$$^{-2}$ M$_{\sun}$ yr$^{-1}$
if we adopt the calibration of Kennicutt (1983), that is,
SFR (M$_{\sun}$ yr$^{-1}$) = L(H$\alpha$)/1.12$\times$10$^{41}$ ergs s$^{-1}$.
For $q_{0}$=0.5, SFR = 135 $h_{75}$$^{-2}$ M$_{\sun}$ yr$^{-1}$.
This assumes an IMF consistent with Salpeter (1955), an upper
mass cut-off of 100 M$_{\sun}$, and 1.1 magnitudes of extinction
at H$\alpha$.  This value for the extinction is consistent with the
reddening derived from the spectral energy distribution (Figure 2).

If we assume that the galaxy seen as Arc A is magnified by lensing
by a factor of 14.5,
then the star-formation rate for Arc A is 20 M$_{\sun}$ yr$^{-1}$
for $q_{0}$=0.1, and 9.3 M$_{\sun}$ yr$^{-1}$ for $q_{0}$=0.5.
This is very typical of the star-formation rates estimated from
the UV continuum luminosity for
Lyman dropout galaxies or Hubble Deep Field galaxies
(Madau et al 1996;  Steidel et al. 1996a)

Moreover, Arc A in the CVF image
appears bright on its eastern edge compared to the
$K'$ image.  The seeing for the IRTF data was
about 0.4 $\arcsec$, compared to a total linear extent of about
5 $\arcsec$ for the arc, so it is well-resolved spatially.
We also show the WFPC2 image of A2218 through the F702W filter from the HST
archive (the WFPC2 data was presented by Kneib et al. 1996).
The bulk of the excess emission appears to be concentrated in
a region along the arc which is fairly faint in the F702W stellar
continuum (rest frame $\lambda$2000$\AA$). Thus the effect of the lensing
is to stretch out what is probably a small, compact galaxy
and allow us to see that star-formation is not happening in
a single, co-eval and co-spatial burst.

We refit the broad-band spectral energy distribution (SED) of Arc A,
which was discussed by E96.  The theoretical curves shown by E96
did not fit the SED very well, and new models have subsequently
become available (Leitherer et al. 1996 and references therein).
We tried several models, chosen to span the important parameters,
and found  a few good fits to the SED: see Figure 2.
Instantaneous single burst models, e.g. those of Bruzual $\&$ Charlot (1996, 
denoted GISSELL in the figures) or
Fioc $\&$ Rocca-Volmerange (1997, denoted RVF in the figures), 
fit better than continuous
star-formation models, and an age of 40 Myrs gives the best fit,
with E(B-V)=0.18, assuming an LMC-like reddening law.  However,
this model predicts a very small equivalent width for H$\alpha$ emission,
W(H$\alpha$)$\sim$ 0.1 $\AA$ (Leitherer $\&$ Heckman 1995),
in clear conflict with the
observations.  For the reddening
law for starbursts given by Calzetti (1997), which one expects to be more
appropriate, the fit is somewhat worse, and the age derived
quite different: 3.8 Myrs, and E(B-V)=0.35 gives the best fit, and
the predicted W(H$\alpha$) $\sim$ 100$\AA$, in better agreement with
the H$\alpha$ observation.

\begin{figure}[ht]
\vbox to2.6in{\rule{0pt}{2.6in}}
\includegraphics{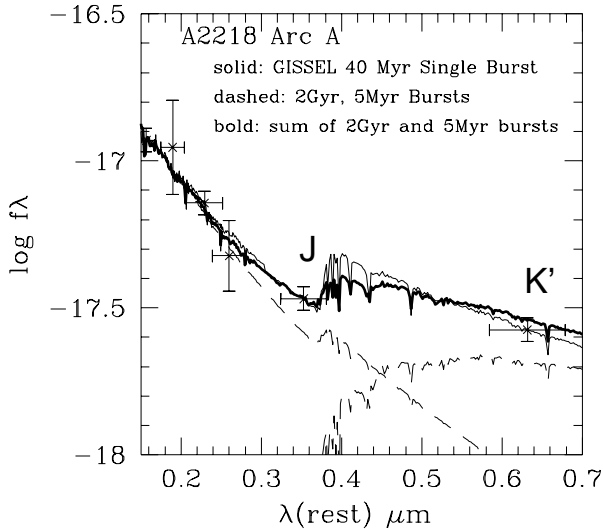}
\caption{Spectral energy distribution for Arc A,
and GISSEL galaxy evolution models, log f$_\lambda$ in cgs units
versus rest wavelength.  The SED is fit by two different
scenarios for the history of star-formation in the galaxy. For details, see text.
}
\end{figure}

Thus the large reddening correction needed in the UV makes conclusions
regarding the age and star-formation rate of Arc A uncertain.
Moreover, the
data points shown in Figure 2 are for the broad band fluxes integrated
over the entire galaxy. 
The H$\alpha$ picture indicates already that the star-formation is not
coeval over the whole galaxy.  An alternative possibility is that star-formation
has been episodic. In Figure 2 we show a second model, where we fit
two bursts, one fixed at an age of 2 Gyr (with LMC reddening; 
the age of the Universe at z$\approx$ 2.5 is not much older than 2 Gyrs) 
and a 5 Myr old burst (and Calzetti reddening).  This two burst model fits
the data very well, and the older burst contains $>$95 percent of the mass
of the galaxy.  This is qualitatively a very different star-formation
history for this galaxy than the single burst model, 
one which the current data cannot rule out.

An older population could be detected by rest-frame near-IR 
photometry, but for z$\sim$2-3, the rest-frame near-IR is redshifted
into the mid-IR, and is beyond the sensitivity of current instruments
for most objects.  In the next section, we describe $ISOCAM$ observations
of MS1512-cB58, at $z=2.7$, which we obtained in order to limit 
the presence of
an older stellar population.

\section{Mid-IR ISOCAM observations of MS1512-cB58}

MS 1512-cB58 is a galaxy which was  discovered serendipitiously
(Yee et al. 1996) 
during the course of the CNOC-1 redshift survey of moderate redshift galaxy 
clusters (Yee, Ellingson $\&$ Carlberg 1996).  It is at $z=$2.7 with apparent 
$V=$20.5.  Its observed optical (rest-frame UV) spectrum 
resembles that of local starburst galaxies (Yee et al. 1996, Steidel
et al. 1996a; see also Conti, Leitherer \& Vacca 1996). 
The lack of a 4000\AA~ break in the rest frame
 (Ellingson et al. 1996) conveniently redshifted so that the broad-band $J$ and $H$ filters
 straddle it, implies a very young age, as does
 the appearance of the C IV $\lambda$1550 P-Cygni profile caused by
 winds associated with the hot stars.
The best fit
to the SED is a continuous star-formation model of about 10-20 Myr,
with LMC reddening of $E(B-V)$ $\sim$ 0.3. 
Bechtold et al. (1997) detected H$\alpha$ emission, confirming 
a high star-formation rate for MS 1512-cB58. 
However, as in the case of the A2218 arc, the limits on the presence
of an older stellar population are weak.  Ellingson et al. (1996) 
showed that  if a two burst model is fit to the spectral energy distribution,
an older, 2 Gyr old burst could comprise as much as 85$\%$ of the galaxy
mass.

MS1512-cB58 has the brightest apparent magnitude of
the high redshift star-forming 
galaxies discovered so far, which is in part a result 
of the relatively young age of its starburst, and in part because of  
magnification by gravitational lensing by a foreground $z=0.37$
cluster of galaxies (Seitz et al. 1998,
Ellingson et al. 1998).  The magnification 
factor is probably as high as 25$\times$.  Thus, MS1512-cB58 is an
ideal candidate for mid-IR observations.

We used the long wavelength camera of the
$CAM$ instrument (Cesarsky et al. 1996) aboard the {\it Infrared
Space Observatory} ({\it ISO}, Kessler et al. 1996) to image MS1512-cB58
through two broad band filters, LW2 (centered on 6.7$\mu$m) and
LW10 (centered on 11.5$\mu$m), covering rest-frame 1-4$\mu$ for
MS1512-cB58.  
Details of the observations and data reduction are given in
Bechtold, Yee $\&$ Ellingson (1998, in preparation).
The data were spatially oversampled in order to resolve MS1512-cB58
from the central cD galaxy of the foreground cluster, which lies only
6 $\arcsec$ away.  The noise in deep $ISOCAM$ images is primarily from
unresolved point sources, and so we evaluated the significance of
all sources from the variation from point to point in the images themselves. 
We require sources to be detected at $>$3$\sigma$ in both filters,
with the result that MS1512-cB58, the central cD and several other
members of the foreground cluster are detected. 

\begin{figure}[ht]
\vbox to2.6in{\rule{0pt}{2.6in}}
\includegraphics{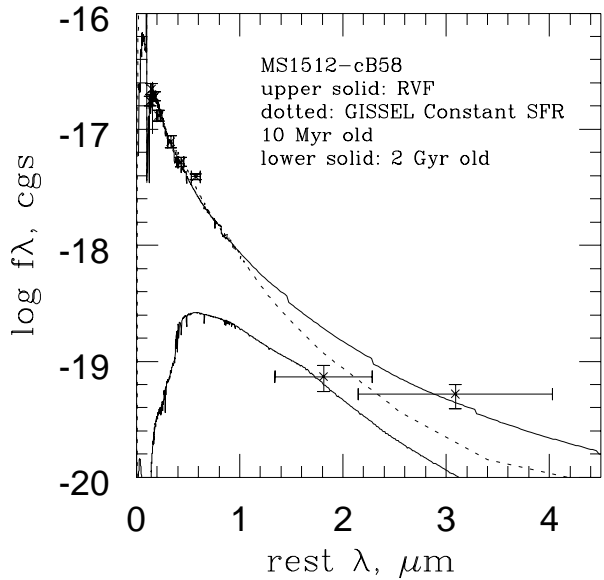}
\caption{Spectral energy distributions of MS 1512-cB58 
with $ISO$ flux points,
and galaxy evolution models, log f$_\lambda$ in cgs units
versus rest wavelength, see text.  
}
\end{figure}

The spectral energy distribution for MS1512-cB58, including the $ISOCAM$
flux points, is shown in Figure 3.  
The $ISOCAM$ flux points fall on the extrapolation of the 
young burst model which best fits the optical and near-IR data points,
with very little room for the addition of an older population.
Less than 10$\%$ of the mass of MS1512-cB58 can be in a 2 GYR
population.  Thus, MS1512-cB58 may well be a true proto-galaxy,
undergoing its very first burst of star-formation.
Futher mid-IR observations of MS1512-cB58 and other high redshift
galaxies will be possible with SIRTF and NGST.

\section{H$\alpha$ from Damped Ly$\alpha$ absorbers}

For many years, quasar absorption lines have been used to study the
evolution of galaxies and the intergalactic medium at high redshift.
The damped Ly$\alpha$ systems in particular 
are generally thought to be the progenitors of present-day galaxies
(Wolfe et al. 1986)
and should thus be sign posts for distant normal galaxies in an early
stage of evolution.  Prochaska $\&$ Wolfe
(1996, 1997) have suggested that the
asymmetric line profiles seen in the metal lines associated with damped
systems support the view that damped Ly-$\alpha$ systems are large rotating
disks.  On the other hand,
Haehnelt, Steinmetz $\&$ Rauch (1998)
suggest that such profiles can occur naturally
in systems composed of smaller protogalactic fragments.
Direct detection of the galaxies causing
damped Lyman-$\alpha$ absorbers will help  
interprete the large body of spectroscopic observations of
them in the context of the formation of galaxies and structure.
Deep imaging of
the fields of damped Lyman-$\alpha$ systems does appear to show a
statistical excess of luminous galaxies near the quasar line of sight
(e.g. Aragon-Salamanca, Ellis $\&$ O'Brien 1996). But without redshifts,
one cannot actually associate any particular galaxy with
a given absorption line system.

We have used the IRTF NSFCAM and Palomar 5m Near-IR CASS camera
to search for H$\alpha$ redshifted into the K band,
associated with damped Ly$\alpha$ quasar
absorption line systems.  Details of the observations are given in
Elston, Bechtold $\&$ Cutri (1998, in preparation).
We obtain images through a 1$\%$ narrow band filter
(CVF) tuned to H$\alpha$ at the redshift of the damped
Ly$\alpha$ absorbers.
So far, this technique has proven very powerful:
we have detected H$\alpha$ companions
for most of the systems observed, with separation of between 2 and 12
arcsec from the quasar line-of-sight.  Often, multiple faint, very compact
objects are found in a single field.   We interprete these  
H$\alpha$ emitters to be clumps of star-forming gas in the same 
sheet or filament which contains the damped Ly$\alpha$ absorber.  

\begin{figure}[ht]
\vbox to2.6in{\rule{0pt}{2.6in}}
\includegraphics{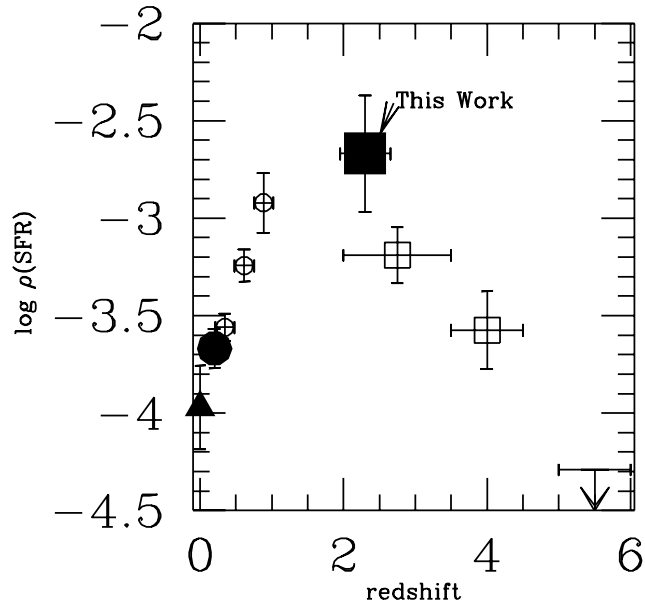}
\caption{
Volume-averaged star formation rate (M$_{\sun}$ yr $^{-1}$ MPC$^{-3}$) as 
a function of redshift.  Open symbols are derived from UV luminosity,
filled symbols from H$\alpha$.
Open circles, Lilly et al 1996; Open squares, Madau et al 1996 (HDF); 
Filled Triangle, Gallego et al. 1996; Filled Circle, Tresse $\&$ Maddox,
1997; Filled square, our work.
}
\end{figure}

We are in the process of confirming these with slit spectra at the
KPNO 4m with CRSP, and UKIRT.  
Using the standard conversion between H$\alpha$ flux
and star-formation rate our typical
3$\sigma$ sensitivity is about 5 M$_{\odot}$ yr$^{-1}$.
If the H$\alpha$ fluxes are attributed to photo-ionization
by young stars it appears that star formation rates of
10-20M$_\odot$ yr$^{-1}$ are pervasive at redshifts of 2.5
in agreement with the Lyman-dropout population found by Steidel et al. (1996)
suggesting they are similar populations.

If these observations of individual systems are representative,
we can combine the star-formation rates 
with the statistics of the incidence of
damped Ly$\alpha$ absorbers in quasar lines-of-sight (Wolfe et al. 1995), to make a 
preliminary estimate of
the volume-averaged star-formation rate (Pei $\&$ Fall 1995; 
Fall, Charlot, $\&$ Pei 1996; 
Madau et al. 1996) at $z\sim$2.5 (not all 
candidates have been spectroscopically confirmed yet).  Figure 4 shows the
result.   Our results are consistent,  
within large uncertainties, with the
results derived from the UV continuum luminosity by Madau et al. (1996).   
However, our data so far suggest that the peak in star-formation
may have taken place at somewhat higher redshift than indicated by
the Madau et al. analysis.
Combining our data with observations of the H$\alpha$ luminosity
of low and moderate redshift galaxies will allow an important check on the 
star-formation rates in the Universe derived from other means. 

\acknowledgments

We are very grateful to the staff of the $IRTF$, $Palomar$  and
KPNO for providing the instrumentation which made this work
possible, and for help at the telescope.
We are indebted to Ken Ganga, Dave Van Buren and the staff at IPAC 
for help with the planning and analysis of the $ISOCAM$ observations.
This work was supported by NSF grants AST-9058510 and AST-9617060,
NASA HST grants AR057850194 and GO074480296, and NASA ISO grant  
NAG5-3359.

%
%

\end{document}